\documentclass[preprint,authoryear,12pt]{elsarticle}

\usepackage{epsfig}
\usepackage{lscape}

\journal{Advances in Space Research}
\newcommand{\aap}{{Astron. Astrophys.}}
\newcommand{\aaps}{{Astron. Astrophys. Suppl.}}
\newcommand{\apj}{{Astrophys. J.}}
\newcommand{\apjl}{{Astrophys. J. Lett.}}

\newcommand{\mnras}{{MNRAS}}

\newcommand{\apss}{{Astrophysics and Space Science}}

\newcommand{\araa}{ARA\&A}

\def\pdot {\dot P}

\def\msun{M_{\odot}}
\def\rsun{R_{\odot}}
\def\lsun{L_{\odot}}
\def\zsun{Z_{\odot}}
\def\mdot {\dot M_{\rm W}}

\def\ltsima{$\; \buildrel < \over \sim \;$}
\def\lsim{\lower.5ex\hbox{\ltsima}}
\def\gtsima{$\; \buildrel > \over \sim \;$}
\def\gsim{\lower.5ex\hbox{\gtsima}}
\def\hd {HD\,49798}
\def\bd     {BD\,+37$^{\circ}$442}
\def\bdd   {BD\,+37$^{\circ}$1977}
\def\bddd {BD\,+28$^{\circ}$4211}
\def\cd     {CD\,--30$^{\circ}$11223}
\def\xmm {\emph{XMM-Newton}}
\def\cha {\emph{Chandra}}
\def\swi {\emph{Swift}}

\begin{document}

\begin{frontmatter}


\title{{\bf X-rays from Hot Subdwarfs} \\ 
}


\author{Sandro Mereghetti and Nicola La Palombara
}
\address{INAF-IASF Milano, v.Bassini 15, I-20133 Milano, Italy }
\ead{sandro@iasf-milano.inaf.it, nicola@iasf-milano.inaf.it}




\begin{abstract}
Thanks to the high sensitivity of the instruments on board the \xmm\  and \cha\ satellites, it has become possible to explore the properties of the  X-ray emission from hot subdwarfs. 
The small but growing sample of  hot subdwarfs detected in X-rays includes binary systems, in which the X-rays result from wind accretion onto a compact companion (white dwarf or neutron star), as well as isolated sdO stars in which X-rays are probably due to shock instabilities in the wind. 
X-ray observations of   these low mass  stars   provide  information which can be  useful also  for our understanding of the winds of more luminous and massive early-type stars and can lead to the discovery of particularly interesting binary systems.
\end{abstract}


\end{frontmatter}

\parindent=0.5 cm

\section{Introduction}

While soft X-ray emission from massive OB stars has been discovered  more than thirty years ago \citep{sew79,har79},  only in recent years, thanks to the great sensitivity of the X-ray instrumentation carried by the \xmm\ and \cha\ satellites, it has become possible to reveal in the X-ray range also hot stars with much smaller masses and luminosities, such as the hot subdwarfs. 
These  stars have  high temperatures, corresponding to O and B spectral types,  but luminosity values that place them below the main sequence in the HR diagram. 

The B type subdwarfs  (sdBs) have masses of $\sim$0.5$\msun$, effective temperatures 20 $<T_{\rm{eff}}<$ 40 kK, and surface gravities  5 $<$ log $g<$ 6. 
They are interpreted as evolved low-mass stars that lost most of their hydrogen envelopes and are now in the He core burning phase. 
Subdwarfs of O spectral type (sdOs) constitute a less homogeneous group compared to the sdBs. 
They show a large range of temperatures (40 $<T_{\rm{eff}}<$ 100 kK) and surface gravities 4 $<$ log $g<$ 7, and 
comprise both He-rich and He-poor stars.

X-ray emission  associated with a hot subdwarf  was seen for the first time in 1979 (see Sect.~\ref{sec_hd}),  but it took almost two decades  to  demonstrate that most of the observed X-rays actually originate from its compact companion star \citep{isr97}.   Since then, little progress has been made in this field,  until the developments of the last few years.

Extensive information  on hot subdwarfs can be found in the excellent review by \citet{heb09}, which, however, does not cover their high-energy properties, having been written before most of the results described below were obtained.  The aim of this paper is to review the X-ray properties of hot subdwarfs and discuss their relevance in the context of our understanding of mass-loss from hot stars. 
After a brief introduction on the   X-ray emission from early type stars\footnote{An exhaustive coverage of this topic is provided
by the other articles of this Special Issue.}  and on  the current knowledge of mass-loss in hot subdwarfs (Sect.~\ref{sec_relevance}), we describe all the available   X-ray  observations of hot subdwarfs in  Sect.~\ref{sec_obs}. 
The interpretation of these observations, in particular    their implications on the stellar wind properties and on the 
subdwarfs with compact companions, are discussed in Sect.~\ref{sec_disc}.

\section{Relevance of X-ray observations of hot subdwarfs}
\label{sec_relevance}

There are two main processes that can lead to the production of  X-rays in  hot subdwarf stars: wind emission and accretion onto a compact companion.  In both cases, X-ray observations can give information on the star properties, although,   strictly speaking, the latter process does not involve emission from the hot subdwarf itself. The detection of accretion-powered emission gives the possibility to discover the hot subdwarfs with white dwarf or neutron star companions, which are predicted from evolutionary calculations, but not easily identified with optical observations.

It is well known that early-type stars can generate X-rays if their stellar winds contain plasma  sufficiently hot  to emit in this energy range. In the following, we will call  this process  ``intrinsic'' X-ray emission.
Extrapolating to lower luminosities the empirical relation between X-ray and bolometric luminosity observed in normal OB stars, $L_{X}/L_{BOL}$ = 10$^{-7\pm1}$   \citep{pal81,naz09},  one can estimate the expected X-ray emission for hot subdwarfs.  This leads to expected X-ray luminosities of the order of 10$^{27-32}$  erg s$^{-1}$ and 10$^{26-29}$  erg s$^{-1}$  for O and B type subdwarfs, respectively. 
It must be remembered, however, that the above average relation has a large scatter.

The second possibility is that  X-rays  are produced by accretion onto a compact companion star, which can be either a white dwarf (WD) or a neutron star  (NS).\footnote{Also  binaries composed of a hot subdwarf and a  black hole can exist, but their formation is less frequent \citep{nel10}.} 
As a first approximation,  the accretion rate can be estimated using the Bondi-Hoyle formalism,\footnote{We assume that the subdwarf radius is smaller than its Roche-lobe; this condition is verified in all the hot subdwarf binaries discussed below.} according to which the accretion radius onto an object of mass $M$ is given by $R_A=2GM/(V_W^2+V_{ORB}^2)$,  where $V_W$ is the wind velocity and $V_{ORB}$ is the orbital velocity. The mass accretion rate $\dot M_A$ is related to the relative velocity  $V_{R}=(V_{W}^2+V_{ORB}^2)^{1/2}$ and to the wind density   by  $\dot M_A = \pi R_A^2 \rho V_{R}$.   The wind density  $\rho$ at the position of the compact object can be estimated as  $\mdot = 4 \pi a^2 \rho V_W$,   where $\mdot$ is the wind mass-loss rate from the subdwarf and $a$ is the orbital separation.  From these relations  one obtains the accretion-powered luminosity of a  star with mass $M$ and radius $R$

$$L_X = \frac{G M}{R} \dot M_A  = \frac{G M}{R}  \left(\frac{R_A}{2a}\right)^2 \frac{V_R}{V_{W}} \mdot  \sim  \frac{G M}{R}  \left(\frac{R_A}{2a}\right)^2  \mdot   ~~~~~~~~(1) $$

Obviously, accretion onto a companion and intrinsic emission are not mutually exclusive, and both processes can occur in binary subdwarfs. In the lack of adequate  X-ray data, it can be  difficult to distinguish between the two possibilities, but,  in both cases, the  X-ray emission  depends on the properties of the subdwarf's  stellar wind. Therefore,  X-ray observations provide a new diagnostic tool to investigate the poorly constrained mass-loss processes occurring in these stars.

\subsection{X-ray emission in hot stars}
\label{sec_xray}

Stars of O and B spectral type are sources of soft X-rays with luminosity up to a few 10$^{33}$ erg s$^{-1}$ and thermal spectra corresponding to plasma temperatures of a few million degrees. 
As well demonstrated, e.g.,  by the case of $\zeta$ Puppis \citep{her13}, high resolution X-ray spectra of the brightest OB stars can provide a wealth of information through the study of emission lines.
However, for the majority of the early type stars detected in X-rays, only low resolution spectra are available, which 
can be adequately fit using simple models of  thermal plasma emission.  
For example, the spectra of a large sample of massive OB stars detected with \xmm\ were described  with either a single Mekal model with kT$\sim$0.2--1  keV, or with the the sum of two or three Mekal models of different temperatures  \citep{naz09}.
These data also showed evidence for additional\footnote{With respect to the value expected for the interstellar medium along the line of sight.} absorption, probably occurring in the stellar wind,  in O stars but not in B stars. 
 
Early-type stars are characterized by winds with typical mass-loss rates $\mdot$ in the range 10$^{-7}$--10$^{-5}$ $\msun$  yr$^{-1}$ and terminal velocities of a few thousands km s$^{-1}$. 
It is believed that the observed  X-rays  are produced in these stellar winds, where the  gas is shock-heated by instabilities \citep[see, e.g.,][for a review]{owo13}. 
The main properties of the winds in OB stars are explained in the context of the radiative line-driven wind theory \citep{cas75,kud00}, according to which part of the radial momentum of the photons emitted from the star is transferred to the wind matter through line absorption and reemission. 
The theory predicts  a dependence of $\mdot$ on the star luminosity approximately  given  by $\mdot$ $\propto L^{\alpha}$, with $\alpha\sim$1.5--2. Since the photon absorption/emission process occurs mainly in the metals present in the wind, the mass-loss rate depends also on the metallicity $Z$, with $\mdot$ $\propto Z^{b}$   and $b\sim0.6-0.7$ \citep{vin01}. These theoretical scaling laws  are generally in  good agreement with  the  observational data. However, some discrepancies have been found in  stars with low-density winds,   which show mass-loss rates one or two orders of magnitude below the predicted values \citep{mar05,mar09}.  Such discrepancies might be, at least partially, explained by the fact that the UV line diagnostics used in these stars  underestimate the actual mass-loss rates. 
The study of hot subdwarfs in the X-ray band, providing alternative handles on the properties of weak winds,  can be of interest in this respect.

\subsection{Stellar winds in hot subdwarfs}
\label{sec_wind}

As mentioned above, the theory of radiatively-driven winds  predicts that, for a given temperature and composition, the mass-loss rate scales  with the star  luminosity.    This leads to the simple expectations that  the winds of hot subdwarfs should be weaker  than those of main sequence, giant and supergiant OB stars  and that sdOs should have stronger winds than sdBs.  

These predictions are indeed confirmed by the observations:  evidence for stellar winds has been reported for some sdOs, but not yet conclusively for any sdB star.  
In fact, the mass-loss rates of a few sdOs have been  derived from the P-Cygni profiles of the 
C$_{IV}$ and N$_{V}$  lines in the UV, yielding values of $\mdot\sim10^{-9}$--10$^{-8}$  $\msun$ yr$^{-1}$ \citep{ham81,jef10,ham10}.
On the other hand, for what concerns sdB stars, the only  evidence for mass-loss comes from the observation
of a few objects with  anomalous profiles of the H$_{\alpha}$ and 
HeI lines, interpreted as possible hints of a weak stellar wind \citep{heb03}.

\citet{vin02}  computed models of radiation-driven winds for low mass stars with  10 kK $< T_{\rm{eff}}<$ 35 kK, obtaining the following   relation between the mass-loss rate and the star mass, $M$, luminosity, $L$, and metallicity $Z$: 
 
$$ {\rm log} \dot {M}_W = - 11.70(\pm0.08) + 1.07(\pm0.32)~ {\rm log} (T_{\rm{eff}}/20~\rm{kK})  $$
$$ + 2.13(\pm0.09)~({\rm log} (L/\lsun) -1.5)   $$
$$ - 1.09(\pm0.05)~{\rm log} (M/0.5 \msun) + 0.97(\pm0.04)~ {\rm log} (Z/\zsun) $$ 

\begin{figure}
\begin{center}
\hspace{-3.cm}
\includegraphics*[width=11.5cm,angle=-90]{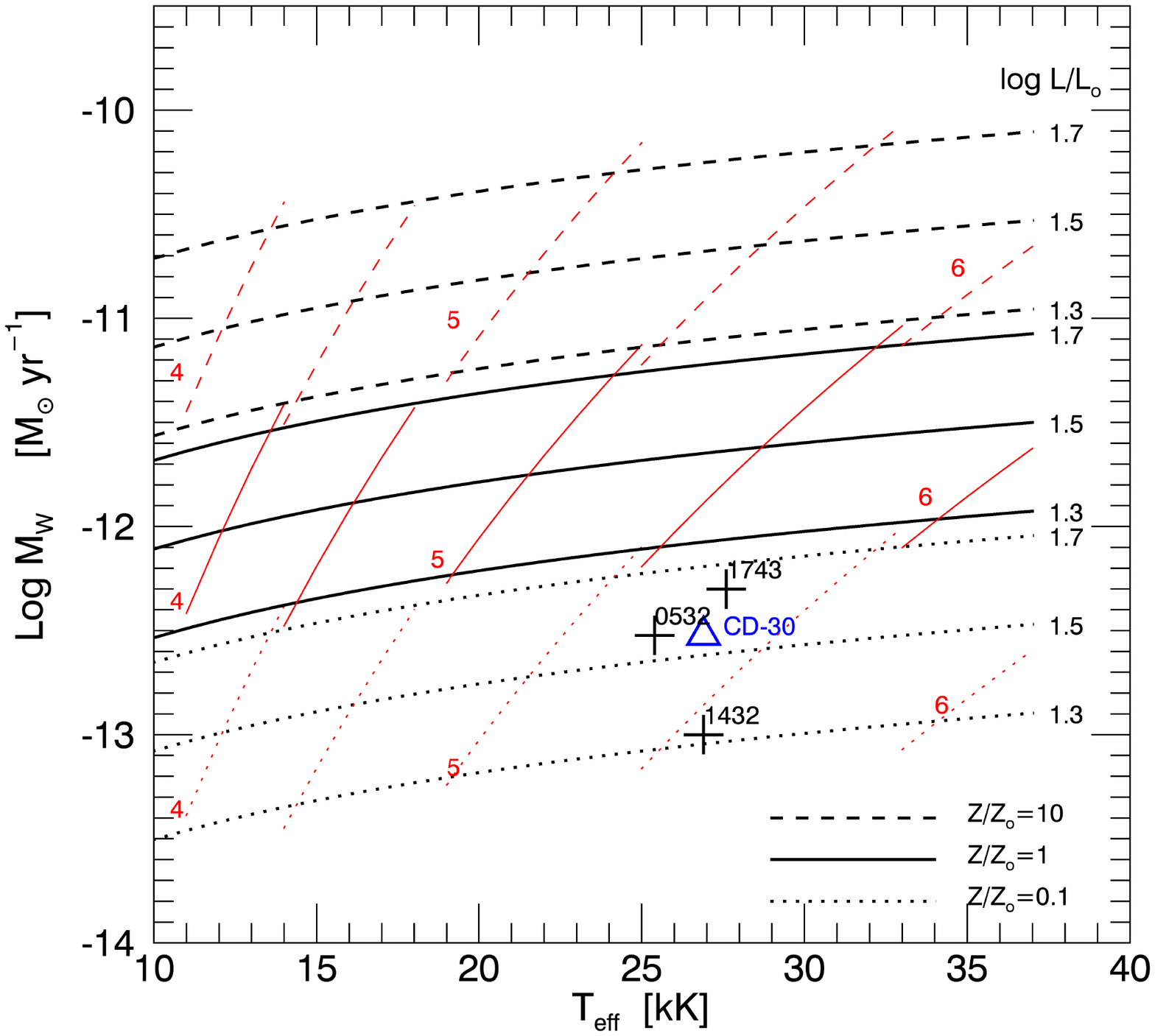}
\end{center}
\caption{Mass-loss rate as a function of effective temperature, according to the relation of  \citet{vin02}, for a star of 0.6 $\msun$. Each set of lines refers to a different metallicity value  ($Z$=0.1 $\zsun$ dotted lines, $Z$=1 $\zsun$ solid lines, $Z$=10 $\zsun$ dashed lines)  and three luminosity values  (log ($L/\lsun$) =1.3, 1.5, 1.7).
The thin red lines show the corresponding values of log $g$.
The triangle indicates the upper limit, derived from X-ray observations, on the mass-loss rate of the sdB star \cd , which has a WD companion.  
The crosses indicate the mass-loss rates upper limits derived for three other sdB stars (PG~1432$+$159, PG~1743$+$477, HE~0532$-$4503), with  the hypothesis that they have neutron star companions (see Sect.~\ref{sec_ul}). 
\label{fig_vc02}}
\end{figure}

\noindent
The range of validity of this relation  (0.5 $<M/\msun<$ 0.7,     1.3 $<$ log$(L/\lsun)<$ 1.7,  and   0.1 $<Z/\zsun<$ 10) is   appropriate for sdB stars.   Fig.~\ref{fig_vc02} shows the mass-loss rates as a function of $T_{\rm{eff}}$,   computed with this relation  for the case of $M=0.6 \msun$  and different values of $L$ and $Z$. 
\citet{ung08} computed mass-loss rates for sdBs obtaining values similar to those of \citet{vin02} for stars of solar metallicity, but much lower values in the case of    $Z<\zsun$ and high surface gravities. 

The wind composition is an important parameter affecting the mass-loss rate, but unfortunately only limited information on the metallicity of hot subdwarfs is available.  The above theoretical estimates for  $\dot {M}_W$ are based on the assumption that the wind behaves as a single fluid, i.e. there is sufficient Coulomb coupling between the metals which are accelerated and the  H and He atoms which constitute the bulk of the wind matter. Such a condition might not be valid in very low-density winds \citep{krt10}. 

\begin{landscape}
\begin{table}
\caption{Hot subdwarfs detected in X-rays}
\begin{tabular}{lccccc}
\hline
        &  HD49798 & BD+37$^{\circ}442$ & BD+37$^{\circ}$1977 &  Feige34 &BD+28$^{\circ}$4211 \\
\hline
Spectral type                &  sdO6   &    sdO9   &    sdO5  &  sdO  &  sdO \\
Luminosity  ($\lsun$)   & $1.4\times10^4$  & $2.5\times10^4$ & $2.5\times10^4$ & 400 & 90 \\
 $T_{\rm{eff}}$ (kK)        &  46.5 & 48 & 48 & 60 &  82\\
 log $g$  (cgs)        & 4.35   & 4 & 4  & 5.2 &  6.3 \\
  $d$   (kpc)           & 0.6   & 2 & 2.7 & 0.3 & 0.1 \\
Parallax (mas)      &  1.2$\pm$0.5 & 3.96$\pm$1.07& --0.5$\pm$1.84 & 3.09$\pm$1.93 & 10.89$\pm$1.46 \\
 $V$                   & 8.3                & 10                      & 10.2                    &    11.2                 &  10.5              \\
 $B-V$               &  --0.27          &  --0.28              & --0.24                &  --0.3                 & --0.34  \\
 $F_X$$^{(a)}$ (erg cm$^{-2}$  s$^{-1}$)  &   6$\times10^{-14}$ $^{(b)}$ &  3$\times10^{-14}$ & 4$\times10^{-14}$  & 3$\times10^{-14}$ &   $10^{-14}$\\
 $L_X$ (erg  s$^{-1}$)         &  2.5$\times10^{30}$  $^{(b)}$ & $10^{31}$  & 3$\times10^{31}$ & 3$\times10^{29}$ & $10^{28}$  \\
 $P$     (s)        &  13.18 & 19.2 ?& -- & --& -- \\
 $P_{ORB}$ (d)  &  1.55  & --  & -- & -- &   --  \\
 $F_X$$^{(a)}$   (erg cm$^{-2}$  s$^{-1}$)  &    7$\times10^{-13}$ $^{(c)}$& -- & -- & -- & -- \\
 $L_X$    (erg  s$^{-1}$)         &  3$\times10^{31}$ $^{(c)}$ &-- &-- & --&  --\\
Log $\mdot$  ($\msun$ yr$^{-1}$) & --9.2 & --8.5    &  --8.5   &  --  &  -- \\
 V$_W$ (km s$^{-1}$)  &  1200  &   2000  &  2000 &  -- & --\\
 \hline
\end{tabular}
\label{tab_det}

\begin{small}
\textbf{Notes:} 

{$^{(a)}$} Observed flux in the  0.2--10 keV range.

$^{(b)}$ During the eclipse of the compact companion

$^{(c)}$  Flux and luminosity of  compact companion

\end{small}

\end{table}
\end{landscape}

\section{X-ray observations of hot subdwarfs}
\label{sec_obs}

The main properties of the five hot subdwarfs that have been detected in the X-ray band are given in Table~\ref{tab_det}. These stars are individually discussed in the following subsections.   In Table~\ref{tab_ul}  we list all the hot subdwarfs that have been observed with sensitive X-ray telescopes (\xmm , \cha , \swi), but without a significant detection.

\begin{figure}
\begin{center}
\hbox{ 
\hspace{-2.5cm}
\includegraphics*[width=9cm,angle=-90]{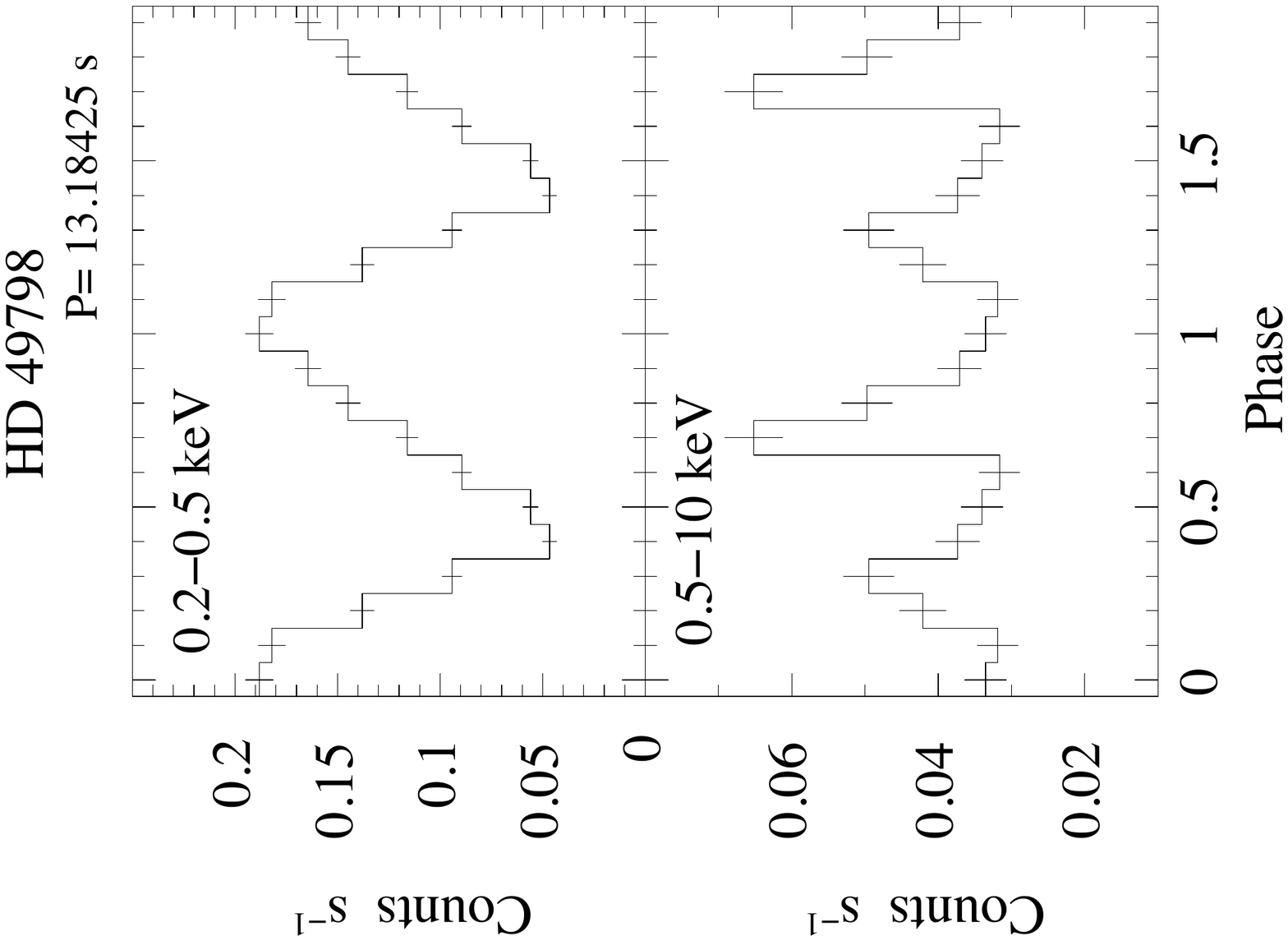}
\hspace{-4cm}
\includegraphics*[width=8.5cm,angle=-90]{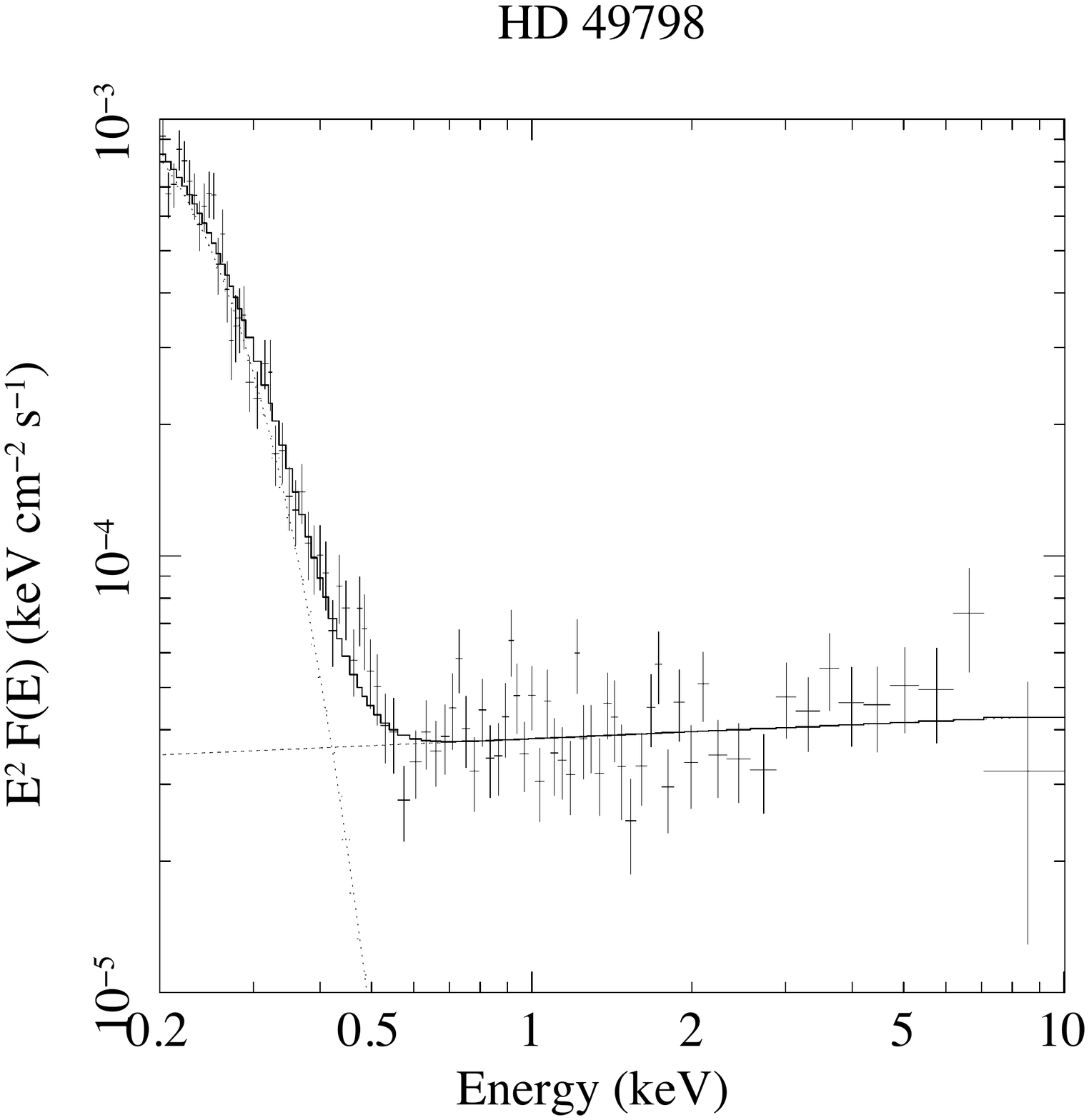}
}
\end{center}
\caption{Results for \hd\ obtained with the $EPIC$ instrument on \xmm .  Left panel:  X-ray light curve in two energy ranges folded at the spin period of 13.2 s. Right panel: X-ray  spectrum   fitted with a blackbody plus power law model (observation performed in October 2014).
\label{fig_sp_hd}
}
\end{figure}

\begin{figure}
\begin{center}
\includegraphics*[width=12cm,angle=0]{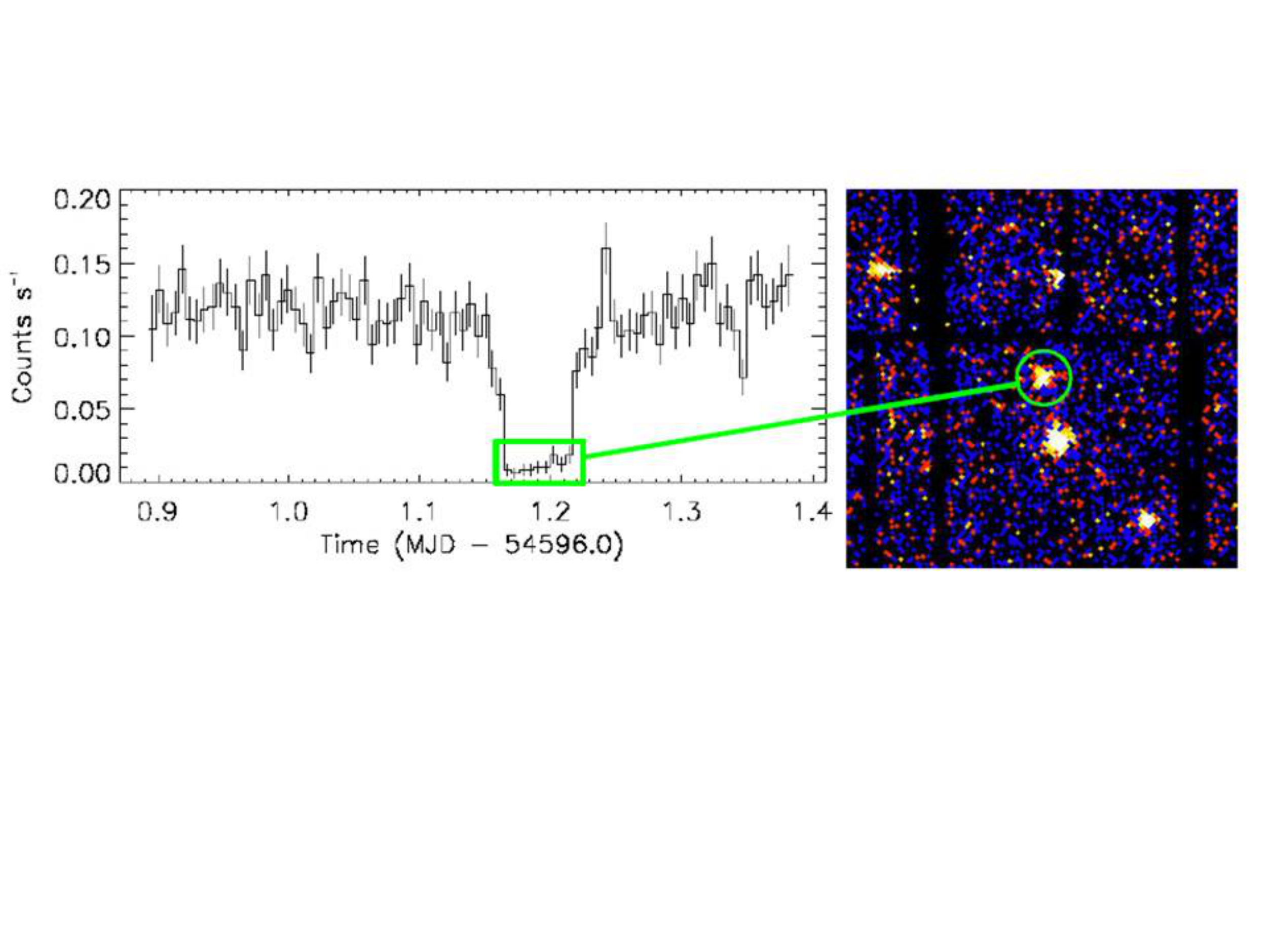}
\vspace{-4cm}
\end{center}
\caption{Left panel: X-ray light curve of   \hd\  obtained  in May 2008 showing the eclipse with duration of 1.3 hours. Right panel: X-ray image (0.2--10 keV)  accumulated during the eclipse in which emission from \hd\ is clearly visible. 
\label{fig_ecl}}
\end{figure}

\subsection{HD\,49798}
\label{sec_hd}

\hd\  is the X-ray brightest hot subdwarf and the only one for which most of the observed emission can  be unambiguoulsy attributed to accretion onto a compact companion star.  Early observations of this bright  (V=8.3), single-lined spectroscopic binary revealed its orbital period  of 1.5477 days and led to the measurement of the optical mass function f(M)= 0.27 $\msun$   \citep{tha70,sti94}, but the nature of its unseen companion remained unknown for many years.
Extensive optical/UV spectroscopic studies indicated that \hd\ is a luminous  ($\sim5\times10^{37}$ erg s$^{-1}$) subdwarf of  O6 spectral type, with  peculiar abundances: it is rich in He and N, while C is underabundant (see Table~\ref{tab_spec}). These abundances are consistent 
with evolutionary models according to which \hd\  is the core of an intially much more massive star which lost its H envelope, most likely during a common-envelope phase. 
 
X-rays from \hd\ were  first detected in 1979 with the {\it Einstein Observatory}, and later shown to be  modulated with a period P=13.2 s  thanks to  a $ROSAT$ observation carried out in 1992 \citep{isr97}.  Such a short and regular periodicity can only be produced by the rotation of either a WD or a NS. As first suggested by \citet{tha70}, the fact that the companion star is a degenerate object explains why it is not visible in the optical/UV  spectra, which are dominated by  the emission of the much larger and brighter sdO star.

\hd\ has been repeatedly observed with the  \xmm\   and \swi\ satellites.  These data, spanning the years from 2002 to 2014,  as well as the few previous observations with other satellites, indicate a remarkably stable X-ray emission. No significant variations were seen in the source flux, spectral shape, pulse profile, and spin period,  with a limit on the period derivative of  $|\pdot |<6\times10^{-15}$ s s$^{-1}$ \citep{mer13}.

A dynamical measurement of the masses of \hd\ and its compact companion has been obtained by the analysis of the orbitally-induced phase delays of the X-ray pulses, combined with the optical mass function and with the system inclination estimated from the duration of the X-ray eclipse \citep{mer09}.  
The measured mass of \hd\  is 1.50$\pm$0.05 $\msun$, among the highest seen in hot subdwarfs, while that of its compact companion is 1.28$\pm$0.05  $\msun$,   consistent with either  a NS or  a massive WD.

Both the timing and spectral properties indicate that the X-ray emission from the compact companion of \hd\   consists of two different components \citep{mer11,mer13}. The spectrum (Fig.~\ref{fig_sp_hd}, right panel) comprises a soft thermal component, well fit by a blackbody of temperature  kT$_{BB}\sim$30 eV and a hard component which dominates the emission above 0.5 keV. The latter is well fit by either a power law of photon index $\Gamma\sim$2 or a thermal bremsstrahlung with temperature  kT$_{BR}\sim4$ keV. 
The folded pulse profile is single-peaked and strongly modulated below 0.5 keV, where the soft thermal component dominates, while it is  double-peaked at higher energies   (see Fig.~\ref{fig_sp_hd}, left panel).
The striking difference in the pulse profile at low and hard energies suggests an interpretation in terms of
two separate physical components, e.g. thermal emission from a hot spot on the star surface and non-thermal
emission with a more complex beam pattern produced in the magnetosphere.  
The  luminosity of L$_{X}=3\times10^{31}$ erg s$^{-1}$ (0.2-10 keV, d=0.6 kpc)
indicates that the accreting companion of \hd\ is most likely a WD\footnote{This is also supported by the emission radius 
derived from the blackbody fit,  $\sim$40 km, larger than that of a NS and consistent with a hot spot on the surface of a WD.}, 
since accretion onto a NS would result in a luminosity larger by a factor 
R$_{WD}$/R$_{NS}$$\sim$300   (see Eq.(1)). 
 
As shown in Fig.~\ref{fig_ecl},  X-rays from the \hd\ binary system are detected also  when the compact object is eclipsed by the sdO star, which has a radius  (1.45$\pm$0.25 $\rsun$,  \citet{kud78})  much larger than that of its companion  ($\sim$3000 km for a massive WD or   $\sim$10 km for a NS). The X-ray  emission during the eclipse is fainter, by a factor $\sim$10, and harder than that seen in the orbital phases outside the eclipse.   Its spectral properties  and likely  origin as intrinsic emission from \hd\  are discussed in Sect.~\ref{sec_disc_intr}.

\subsection{BD\,+37$^{\circ}$442}

\bd\ is a luminous ($\sim10^{38}$ erg s$^{-1}$), extremely helium-rich sdO star \citep{reb66,hus87}.  Its temperature, luminosity, and surface gravity, as well as its mass-loss properties  \citep{jef10},  are similar to those of \hd\ (see Table~\ref{tab_det}).
However, contrary to \hd , no evidence for it being member of a binary system has been reported in the literature. 
A recent spectroscopic search for a binary signature did not detect any radial velocity variation on timescales from hours to months, down to a level of  a few km s$^{-1}$ \citep{heb14}. 

Motivated by  the X-ray detection of \hd\ during eclipse, we  searched  for intrinsic X-ray emission 
from this allegedly single sdO star  with  an  \xmm\ observation  carried out in  August 2011. 
Soft X-ray emission from \bd\  was clearly detected, with  a flux of $3\times10^{-14}$  erg cm$^{-2}$ s$^{-1}$ (0.2-1 keV). Quite unexpectedly, a timing analysis of the X-ray data revealed (at  $\sim3\sigma$ statistical significance) a periodicitiy at 19.2 s  \citep{lap12}.
Fitting the X-ray spectrum of \bd\ with the sum  of  a soft blackbody and a   power law component yields parameters similar to those of \hd , but with larger uncertainties:
fixing the  photon index to $\Gamma$=1 or 3, blackbody temperatures in the range $\sim$30--68  eV are obtained.  The poorly constrained  blackbody temperature and normalization do not permit a precise estimate of the  total X-ray luminosity, which could be in the range from $\sim10^{32}$ to $10^{35}$ erg s$^{-1}$.   
If confirmed, the  periodicity at 19.2 s detected in \bd\ requires the presence of a compact companion, which, considering the large uncertainty on the X-ray luminosity, could be either a WD or a NS.  

\begin{figure}
\begin{center}
\hbox{ 
\hspace{-1.5cm}
\includegraphics*[width=8cm,angle=-90]{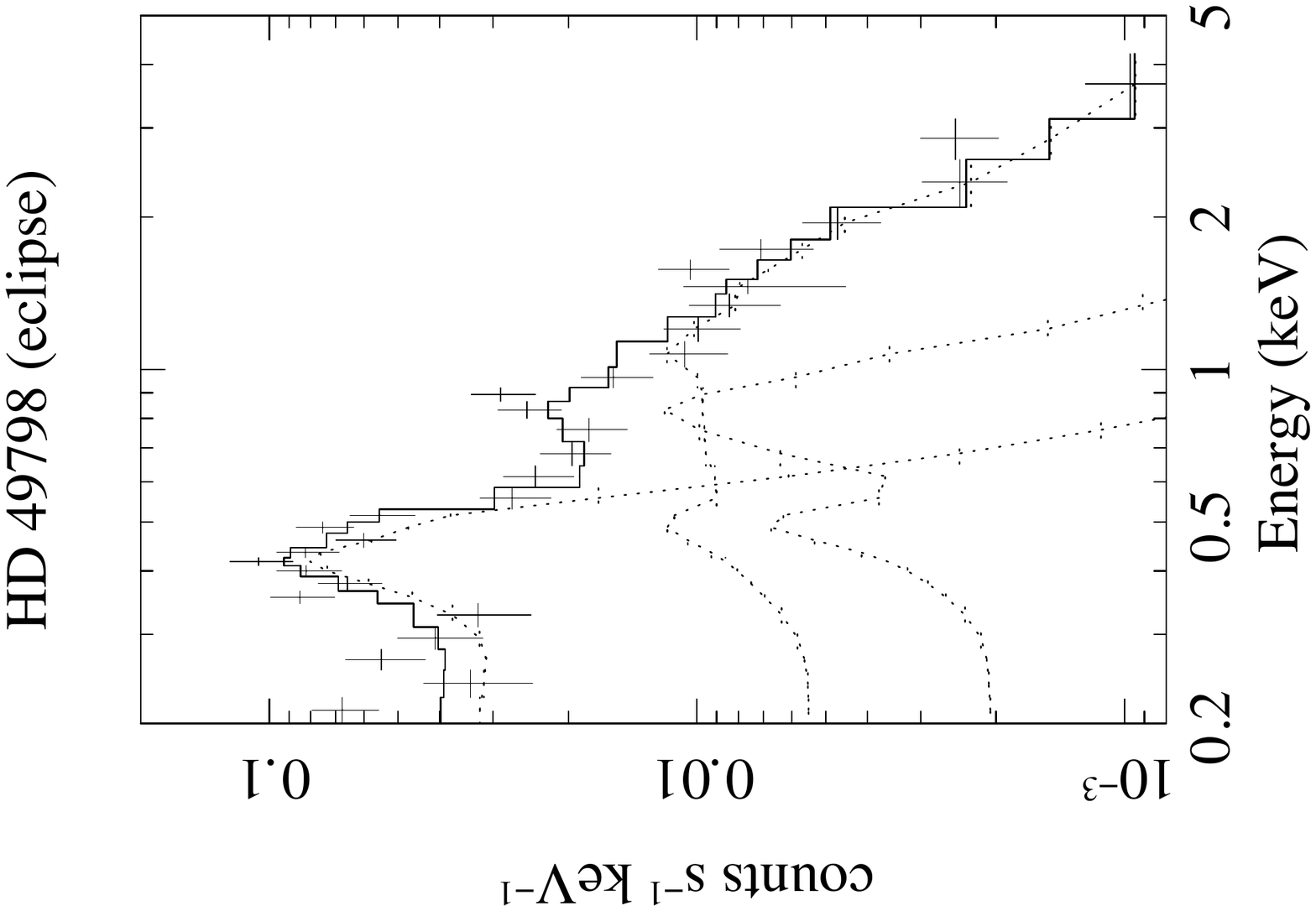}
\hspace{-6.5cm}
\includegraphics*[width=8cm,angle=-90]{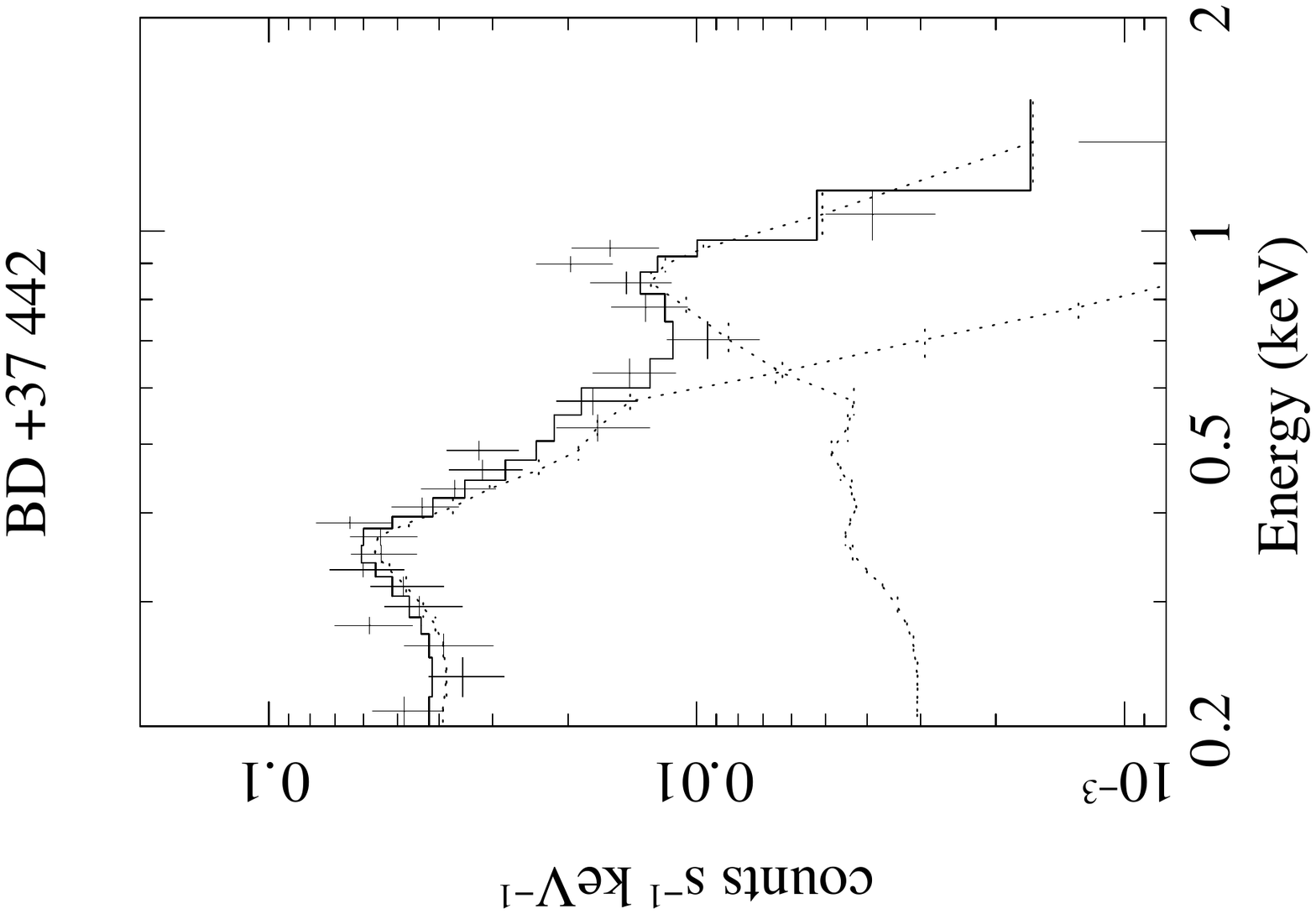}
\hspace{-7cm}
\includegraphics*[width=8cm,angle=-90]{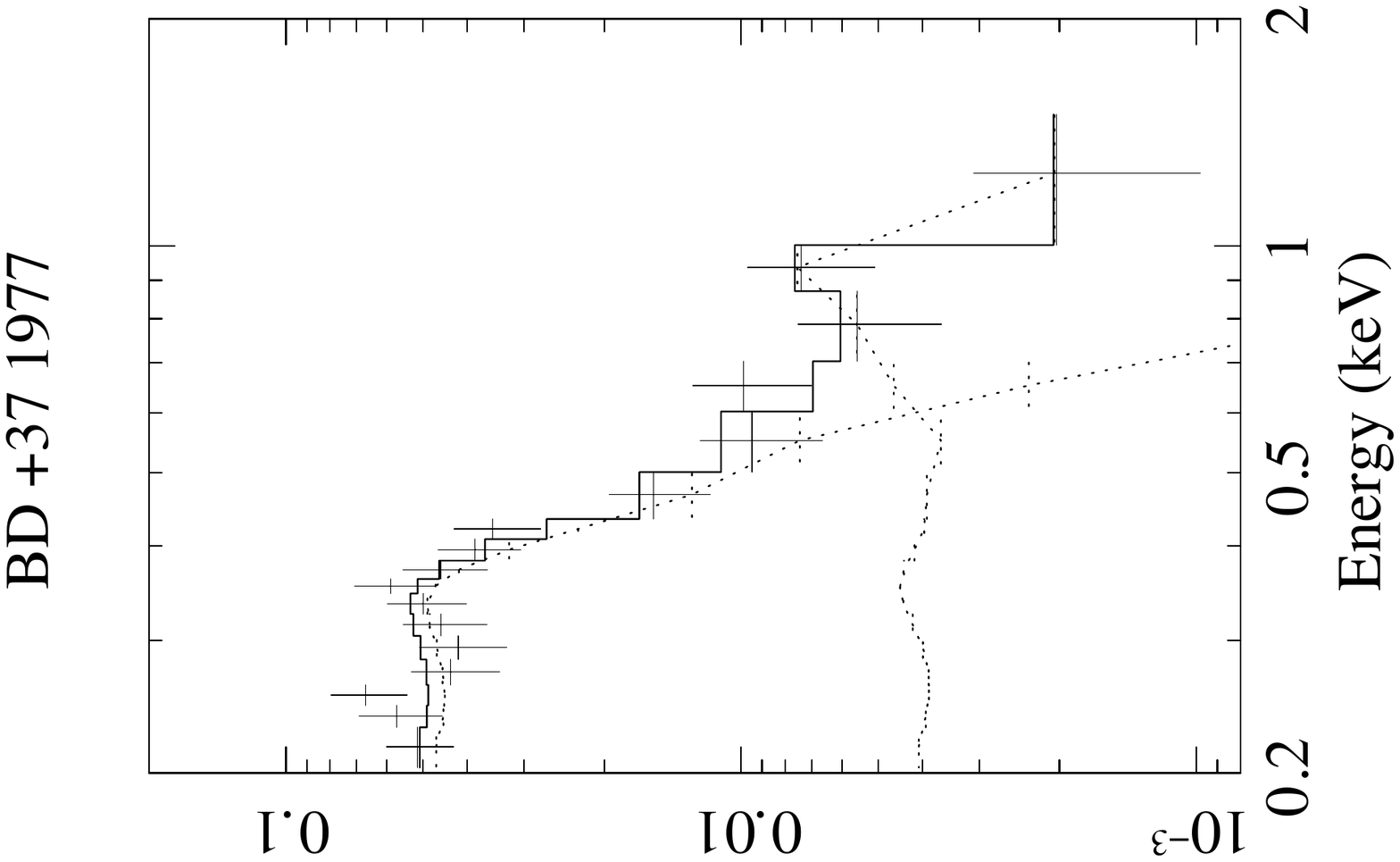}
}
\end{center}
\caption{Left panel: \xmm\ EPIC (MOS+pn) spectrum of  \hd\ during the eclipse. The solid line shows the best  fit with  three thermal plasma models (Mekal) with  temperatures  kT$_1$=0.11 keV, kT$_2$=0.57 keV, and kT$_3$=4  keV (fixed).
The spectrum has been obtained by summing observations performed in 2008, 2011 and 2013, giving a total exposure time of 25 ks.    
Middle panel: \xmm\ EPIC (MOS+pn) spectrum of \bd\ fitted with two Mekal models  with  temperatures kT$_1$=0.11 keV and kT$_2$=0.65 keV.  
Right panel: \xmm\  pn spectrum of \bdd\ fitted with two Mekal models  with  temperatures kT$_1$=0.13 keV and kT$_2$=0.79 keV. 
In all the spectra  the interstellar absorption and the abundances have been fixed at the values given in Table~\ref{tab_spec}. The dashed lines show the individual Mekal components. 
\label{fig_mekal}}
\end{figure}

If \bd\ is indeed in a binary system, the lack of radial velocity variations in its spectrum is puzzling. 
Three explanations are possible: a) a very small orbital inclination; b) a very long orbital period ($>$ several months);  c) the reported pulsations are not real.  Case a) seems  unlikely because the large projected rotational velocity (60 km s$^{-1}$, \citet{heb14}) would imply a significant misalignment between the orbital and the star rotation axis.  Also case b) presents some difficulties because a
large orbital separation would probably result in a low mass accretion rate, unless the orbit is highly eccentric, causing X-ray emission  mainly close to periastron passage (in this case time variability of the X-ray emission is expected).   
More observations are needed to confirm the presence of pulsations and investigate more deeply the different  possibilities. 
As discussed in Sect.~\ref{sec_disc_intr}, the X-ray spectrum of \bd\ can also be fit by a sum of thermal plasma models (see Table~\ref{tab_spec} and Fig.~\ref{fig_mekal}, middle panel),  consistent with intrinsic emission from an isolated sdO.

\begin{table}
\caption{Upper limits on the X-ray emission from hot subdwarfs}
\begin{tabular}{lccccc}
\hline
Name               & Type                    & $d$   &  $P_{\rm ORB}$ & $L_X$$^{(a)}$         & Ref.$^{(b)}$ \\
                       &                          & (kpc)  & (days)         & (erg s$^{-1}$) &   \\
\hline
\cd\                 & sdB+WD            & 0.36 & 0.049 &  $<$1.5$\times$10$^{29}$ & 2\\
PG 1043+760  & sdB+WD?           & 0.12 & 0.66  &  $<$6.8$\times$10$^{30}$ & 1\\
PG 1432+159  & sdB+NS/BH?      & 0.22 & 0.8     & $<$9.9$\times$10$^{30}$ & 1 \\
PG 2345+318  & sdB+WD?           & 0.24 & 0.9    & $<$1.3$\times$10$^{31}$  & 1\\
HE 0532-4503&sdB+NS/BH?        & 0.27 & 2.8    & $<$7.4$\times$10$^{31}$  & 1\\
CPD --64 481  & sdB+WD?            & 0.28 & 0.21   & $<$6.8$\times$10$^{29}$ & 1  \\
PG 1101+249 &sdB+WD/NS/BH? & 0.35 & 0.39  & $<$1.7$\times$10$^{30}$  & 1\\
PG 1232--136 & sdB+BH?             & 0.363 & 0.57  &  $<$5.0$\times$10$^{29}$  & 2\\
GD 687          & sdB+WD?            & 0.38 & 1.1     & $<$1.0$\times$10$^{31}$ & 1 \\
HE 0929--0424&sdB+WD/NS/BH? & 0.44 & 1.9    & $<$2.9$\times$10$^{31}$  & 1 \\
PG 1743+477 & sdB+NS/BH?      & 0.52  & 1        & $<$1.8$\times$10$^{31}$ & 1 \\
PG 0101+039 & sdB+WD?           & 0.57  & 0.33  & $<$1.0$\times$10$^{30}$  & 1 \\
TON S 183     & sdB+WD?           & 0.83   & 0.54  &$<$2.5$\times$10$^{30}$  & 1\\
BD+75$^\circ$325    & sdO  & 0.136 & - & $<$5.8$\times10^{28}$   & 3 \\
BD+25$^\circ$4655  & sdO  & 0.11 & - & $<$5.1$\times10^{28}$   & 3 	\\
BD--22$^\circ$3804 & sdO  & 0.185 & - & $<$9.7$\times10^{28}$   & 3 \\

BD+39$^\circ$3226  & sdO  & 0.235 & - & $<$1.5$\times10^{29}$   & 3 \\
BD--03$^\circ$2179 & sdO  & 0.631 & - & $<$2.4$\times10^{30}$   & 3  \\

CD--31$^\circ$4800 & sdO & 0.132 & - & $<$5.1$\times10^{28}$   & 3  \\
BD+48$^\circ$1777 & sdO   & 0.163 & - & $<$1.2$\times10^{29}$   & 3\\
LS V +22$^\circ$ 38   & sdO   & 0.18 & - & $<$1.5$\times10^{29}$   & 3 	\\
LS IV --12$^\circ$ 1  & sdO	& 0.4 	& - & $<$4.4$\times10^{29}$   & 3 	 \\

LSE 153                    & sdO	& 0.25 & - & $<$3.6$\times10^{29}$   & 3 \\
LSS 1275                  & sdO   & $<$1 & - & $<$2.1$\times10^{31}$   & 3  \\
LSE 263                   & sdO	& 0.25 	& - & $<$4.6$\times10^{29}$   & 3 	 \\
BD+18$^\circ$2647& sdO	& 0.275 & - & $<$3.3$\times10^{29}$   & 3  \\
LSE 21                     & sdO	& 0.05 & - & $<$7.1$\times10^{27}$   & 3 \\
LS IV +10$^\circ$ 9 & sdO     & 0.23 & - & $<$1.5$\times10^{29}$   & 3 \\ 
LS I +63$^\circ$ 198 & sdO	& 0.2 	& - & $<$1.4$\times10^{29}$   & 3 	\\

\hline

\hline
\end{tabular}
\label{tab_ul}

\textbf{Notes:} 

{$^{(a)}$} Luminosity in the 0.2--10 keV range, assuming a thermal spectrum similar to that of \bd .

{$^{(b)}$}  (1) \citet{mer11b}; (2) \citet{mer14}; (3) \citet{lap14};
\end{table}

\begin{table}
\caption{Spectral parameters of three sdO stars fitted with the 
sum of two or three thermal spectra.}
\begin{tabular}{lccc}
\hline
         &  HD 49798 & BD+37$^{\circ}442$ & BD+37$^{\circ}$1977  \\
\hline
N$_H$  (cm$^{-2}$)   &  $5\times10^{20}$  &  $5\times10^{20}$  & 10$^{20}$\\ 
kT$_1$  (keV)  &   0.11 & 0.11    &  0.13\\
kT$_2$  (keV)  &    0.57&  0.65  & 0.79 \\
kT$_3$  (keV)  &   4.   &  --  & --\\
\hline
 & & & \\
Abundances$^{(a)}$ & & & \\
$X_{\rm H}$            &  0.19       &     0.0013  &    0.0013   \\
$X_{\rm He}$          &  0.78       &     0.96     &  0.96   \\
$X_{\rm  C}$            &  0.0001   &     0.025    & 0.025    \\
$X_{\rm  N}$            &  0.025    &      0.003    & 0.003    \\
$X_{\rm  O}$            &  0.0028  &      0.005    & 0.005     \\
$X_{\rm  Si}$            &  0.001   &       0.0008 & 0.0008     \\
$X_{\rm  Fe}$           &  0.0011  &      0.0006 & 0.0006    \\
    \hline
\end{tabular}
\label{tab_spec}

$^{(a)}$    mass fraction. 
\end{table}
    
\subsection{BD\,+37$^{\circ}$1977}

This luminous and He-rich sdO has been  dubbed the spectroscopic twin of \bd , because the optical/UV spectral properties of these two stars are very similar. 
The X-ray emission from \bdd\ was discovered thanks to  a \cha\  survey of a flux-limited sample of sdOs \citep{lap14}  and  recently studied in more detail with a dedicated \xmm\ observation in which it was detected in the 0.15-1.5 keV range \citep{lap15}. Its very soft X-ray  spectrum is well fit by the sum of two thermal plasma models with temperatures of about 0.13 keV and 0.8 keV, provided that adequate elemental abundances are used  (see details in Table~\ref{tab_spec}).
The X-ray   luminosity is $3.3\times10^{31}$ erg s$^{-1}$    (for d=2.7 kpc). 
A search for periodicities gave negative results, but, given the faint X-ray flux, the upper limits on the pulsed fraction  are not particularly constraining.   \bdd\ is a bright and relatively well studied star in the optical/UV bands and no radial velocity variations have been reported in the literature. It is thus natural to interpret these results in terms of intrinsic X-ray  emission from the sdO star. 

\subsection{BD\,+28$^{\circ}$4211 and Feige 34}

\bddd\ and Feige 34  are the only other sdOs detected in the \cha\  survey carried out by  \citet{lap14}.  Being very faint sources (X-ray fluxes of a few 10$^{-14}$ erg s$^{-1}$), they yielded  only  a few counts in these short observations:  enough for a significant detection in the low-background \cha\ HRC-I instrument, but insufficient for a  timing analysis.

The parallax of  \bddd\ obtained with  {\it Hipparcos} implies  a  very small distance. Therefore, this star, with an X-ray   luminosity of only  $\sim10^{28}$ erg s$^{-1}$, is  the closest hot subdwarf detected in the X-ray band.
Note that the angular resolution of the \cha\ data is sufficiently good to unambiguosly associate the observed X-ray emission to \bddd\ and not to a  faint companion star lying at an angular distance of only 2.8 arcsec \citep{mas90}. 
A detailed analysis of the UV and optical spectra of \bddd\  with NLTE models indicates a surface gravity in the range 6.1$<$log $g<$6.5 \citep{lat13,lat15}.  Thus, contrary to the three objects discussed in the previous subsections, this star belongs to the class of low-luminosity, compact sdOs. 

The same might be true for Feige 34, for which \citet{the95} derived log $g$=6.8  (this is, however, at variance with the value log $g$=5.2 reported by \citet{wer98}).  Feige 34 has an infrared excess which might result either from free-free emission in a wind or from a late type companion star. The latter possibility is favored by the analysis of \citet{the95}, who estimate that the putative companion should be of M2 spectral type.   Since also late-type stars are known X-ray emitters, it cannot be excluded that (part of) the observed X-ray emission is due to the putative companion. 

\subsection{sdBs with (candidate) compact companions}
\label{sec_ulsdb}

Up to now, no  sdB star has been detected at X-ray energies. 
The recent observations carried out with sensitive satellites focussed on sdBs in binaries  with either confirmed or candidate compact companions. Although none of the observed targets were detected, these observations are relevant because they provide upper limits that are orders of magnitude smaller than those previously available from the {\it ROSAT} All Sky Survey.

\cd\ is the most interesting sdB for which a deep  upper limit on the X-ray luminosity has been obtained. Contrary to the case of the other sdB binaries discussed below, the presence of a compact companion is certain:
it is a WD  with a dynamically-measured mass of 0.7 $\msun$, orbiting the subdwarf star with a period of only 1.2 hr, the shortest known among sdB binaries  \citep{ven12,gei13}.
A 50 ks long observation with \xmm\ yielded a luminosity upper limit  of L$_X\sim1.5\times10^{29}$ erg s$^{-1}$ (for d=364 pc), which, as discussed in Sect.~\ref{sec_ul}, provides interesting constraints on the mass-loss rate from this sdB \citep{mer14}. 
 
The presence of a compact companion has been suggested for several other sdB stars.
These systems with ``candidate'' compact companions have been selected by means of radial and rotational velocity measurements of single-lined spectroscopic binaries \citep{gei10}. The assumption of synchronous rotation in these systems allows one to estimate the orbit inclination and thus to set a lower limit to the mass of the companion. If this limit exceeds the value expected for a late-type main-sequence star, it is likely that the companion is a WD or a NS. 
A search for X-ray emission from a sample of twelve such systems, chosen among those with the lowest interstellar absorption and shortest orbital periods, was carried out with the \swi\ satellite  \citep{mer11b}.  Upper limits in the range $\sim10^{30}-10^{31}$ erg s$^{-1}$ were obtained for their X-ray luminosities (Table~\ref{tab_ul}). The implications for the wind mass-loss rates of these sdBs are discussed in Sect.~\ref{sec_ul}.

\begin{figure}
\hspace{-4cm}
\includegraphics*[width=14cm,angle=-90]{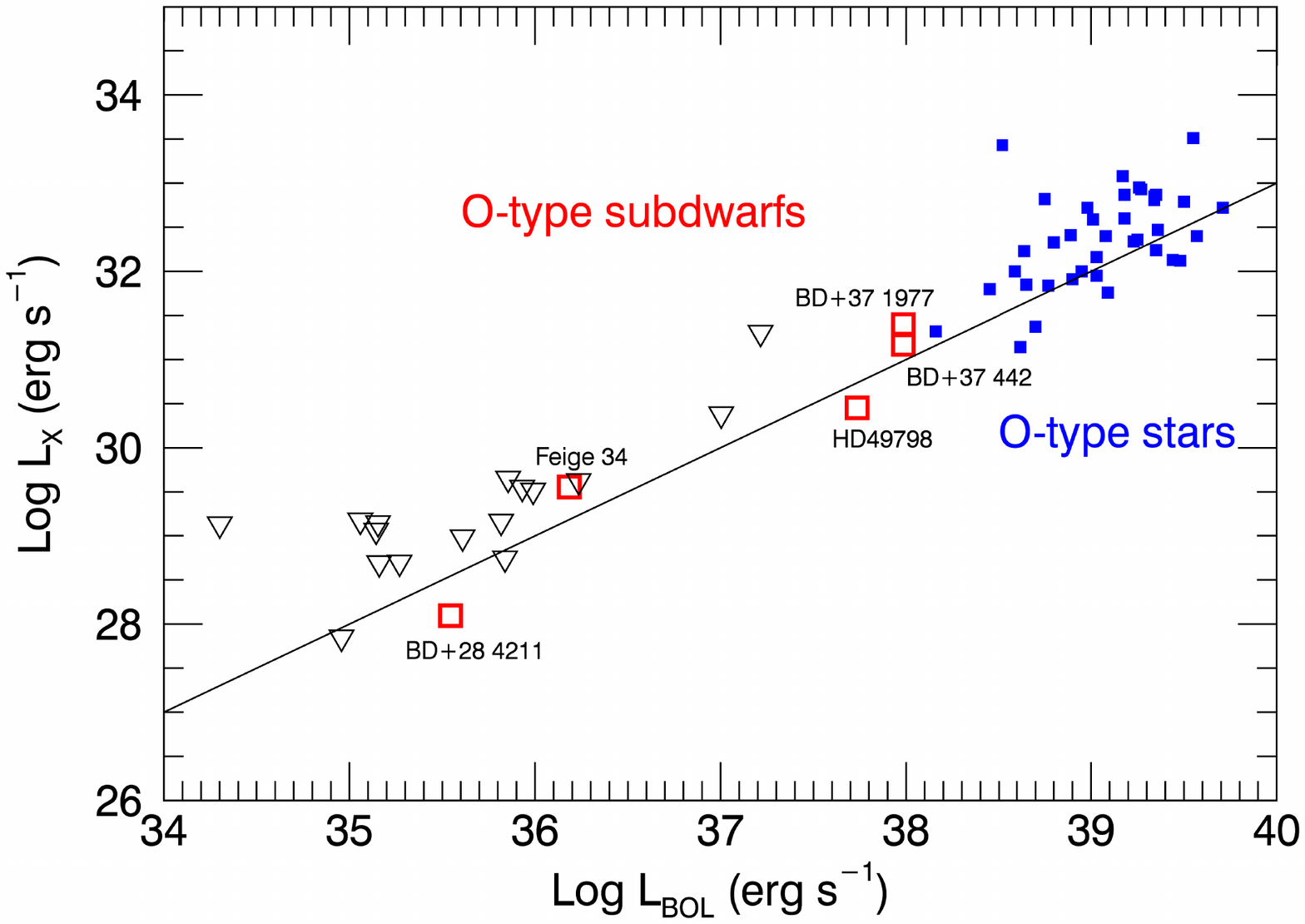}
\caption{X-ray versus bolometric luminosity for hot subdwarfs. The five red squares indicate the sdO detected in X-rays, while the upper limits for the non-detected  sdOs are shown by the black triangles. The blue squares show for comparison  the O stars detected in the ROSAT All Sky Survey \citep{ber96}.  The line indicates the relation $L_{X}/L_{BOL}$ = 10$^{-7}$. 
\label{fig_lxlbol}}
\end{figure}
 
\section{Discussion}
\label{sec_disc}

\subsection{Intrinsic X-ray  emission from hot subdwarfs}
\label{sec_disc_intr}

In Fig.~\ref{fig_lxlbol}  we compare the X-ray and bolometric luminosities of the sdO stars  to those of the  normal O-type stars seen in the {\it ROSAT} All Sky Survey \citep{ber96}.  
For the case of \hd\ we consider here only the X-rays  observed  during the eclipse of its compact companion, which are most likely due to intrinsic emission from the sdO star\footnote{Actually it cannot be completely excluded that  (some of) the X-rays detected  during eclipse are due to emission from the compact object reprocessed in the wind of the hot subdwarf (see discussion in \citet{mer13}); this might explain its harder spectrum  compared to the other sdO stars.}. 
It is clear that the luminosities of the five sdOs detected in X-rays, as well as the upper limits of the undetected ones, are consistent with an extrapolation of the average relation  $L_{X}/L_{BOL}$ = 10$^{-7\pm1}$    followed by the more luminous O-type stars \citep{pal81,naz09}. 

From the X-ray spectral point of view, the sdOs are similar to the normal O-type stars.  
As shown in Fig.~\ref{fig_mekal}, good fits to the X-ray spectra of \hd\  (during the eclipse), 
\bd\ and \bdd\  can be obtained with the sum of two or three plasma emission models with  different temperatures, provided that the proper element abundances are adopted.  
It is interesting to note that these three X-ray emitting sdOs  are among the few hot subdwarfs for which evidence of  mass-loss has been reported. 
No spectral information is available for the two remaining faint sdOs,  which were only observed with the \cha\ HRC instrument.

These findings suggest that the process of X-ray emission due to wind instabilities and shock heating is not restricted to the dense winds of the  luminous and massive hot stars, but it can also operate in the weak winds of sdO stars. 
The five sdOs detected in  X-rays have rather different atmospheric compositions:
\bd\ and \bdd\  have almost pure He atmospheres and are C-rich \citep{bau95,jef10};  \hd\ has one atom of He for each of H and  is N-rich \citep{kud78};  
the compact sdO \bddd\  has solar He  and subsolar to solar abundances of several metals \citep{lat13,lat15};
finally, He is slightly deficient for Feige 34, while Fe and Ni are enhanced by factors of 10 and 70 \citep{wer98}. 
These characteristics can in principle cause differences in their winds which might be explored with future, more sensitive X-ray observations.

\subsection{X-ray emission from compact companions: the case of \hd\ }
\label{sec_disc_accr}

While the possible presence of a compact companion in \bd\ needs a confirmation of the reported 19 s periodicity, there is no doubt that the spectroscopic binary \hd\ hosts either a WD or a NS spinning with a period of  13.2 s.   
As discussed below, accretion onto this compact object is the most natural explanation for the  observed pulsed X-rays.  The presence of a power-law spectral component rules out a purely thermal 
emission powered by the internal heat of a cooling NS, while the possibility of rotation-powered emission seems unlikely. 
The X-ray luminosity of a rotation-powered NS or WD with spin period P=2$\pi$/$\Omega$ is 
$L_X$ = $\eta_X$I$\Omega$$\dot\Omega$, where $\eta_X$ is the efficiency of conversion of  rotational energy into radiation and $I$ is the moment of inertia.
If the \hd\ companion is  a NS ($I_{\rm NS}\sim10^{45}$ g cm$^2$), the minimum spin-down rate required to power the observed luminosity  is $\pdot_{\rm min} = 6\times10^{-15} (L_{X}/10^{32} {\rm erg/s}) (0.001/\eta)$ s s$^{-1}$, barely compatible with  the current upper limit  $|\pdot|<6\times10^{-15}$ s s$^{-1}$ \citep{mer13}. Thus rotation-powered emission from a NS can be ruled out, unless an implausibly high efficiency is invoked. In the case of a WD companion ($I_{\rm WD}\sim10^{50}$ g cm$^2$), the required $\pdot_{\rm min}$, being  a factor $\sim10^5$ smaller,  would be compatible with the observations.
A strongly magnetized WD could 
accelerate relativistic particles, likewise   a radio pulsar, thus producing non-thermal magnetospheric X-ray emission.   Although the existence of rotation-powered ``WD-pulsars''  has been proposed \citep[e.g.][]{uso88}, none  have been found so far.
Also considering that the energetically-dominant thermal component in the \hd\ X-ray spectrum would be difficult to explain in this scenario, we conclude that the possibility of rotation-powered emission  is  quite unlikely and concentrate in the following on the accretion scenario.

Independent on the WD or NS nature of the    companion star, \hd\ does not fill its Roche-lobe. 
Therefore, accretion in this binary occurs by capture of the sdO stellar wind and the expected X-ray 
luminosity can be estimated with Eq.~(1). 
\citet{mer09},  based on  the   wind parameters of \hd\ available at that time  
\citep{ham81},  
concluded that the compact object is a WD, since an accreting NS would have been more luminous than observed.  
An updated estimate of the mass-loss rate (Log $\mdot$ = --9.2  yr$^{-1}$) 
and wind terminal velocity ($V_W$=1200 km s$^{-1}$) has been recently obtained 
(W. Hamann, private communication,  for d=0.6 kpc).
With these values  and  a radius of 3000 km, appropriate for a massive WD, eq. (1)  gives a luminosity of $L_X\sim10^{31}$  erg s$^{-1}$, which is in reasonable agreement with the observed value of
$L_{\rm 0.2-10 keV} = 3\times10^{31} \left(\frac{d}{0.6 kpc}\right)^2$ erg s$^{-1}$.
For the alternative possibility of an accreting NS (R=10 km), eq. (1)  would give  $L_X\sim3\times10^{33}$  erg s$^{-1}$.
Even taking into account that the total luminosity, including the unobserved 
contribution of the blackbody component below 0.2 keV,  could be a factor $\sim$4 higher than the above value,
it seems difficult  to reconcile the expected NS luminosity with the observations, unless the source distance
is  greatly underestimated. 

\citet{ibe94} discussed an evolutionary path leading to a system like \hd . This involves an initial binary
with  masses of $\sim$8--9 $\msun$ in which the WD, formed from the originally
more massive star, is engulfed in the  envelope of the secondary, when the
latter expanded to become a red-giant with a non degenerate He core. 
This  leads to the expulsion of the common envelope  and to the emergence of a much more compact
binary, composed of a CO or ONe WD and a He-star, as currently observed.
\hd\ is now  within its Roche-lobe, but it will expand again in the future,  transferring He-rich matter
onto  its  companion at a higher rate. If the companion is indeed a massive WD, this could lead to a
type Ia supernova explosion or to the formation of a rapidly spinning NS through accretion-induced collapse \citep{wan12,hur10,tau13}.
Being the descendent of intermediate mass stars ($\sim$8--9 $\msun$), the \hd\ binary  would be the progenitor
of a type Ia supernova  with a short delay time, unless, as suggested by \citet{dis11}, the explosion is delayed due to
the fast rotation of the WD. 
In the alternative case of accretion-induced collapse,  the resulting NS would probably have a high
rotational velocity. 
This would be an evolutionary path to create millisecond
pulsars without going through recycling in a low mass X-ray binary \citep{bha91}.
Recently,  \citet{liu15}  performed  population synthesis  computations of different evolutionary channels leading to the 
formation of ONe WD + He-star binaries. They found  that systems with  long orbital period and high He-star mass, as observed  
in \hd , are rarely formed and conclude that the companion of \hd\ is more likely to be a CO WD, 
which can reach Chandrasekhar mass limit and produce a   type Ia SN  after a few 10$^4$ years of mass transfer, as computed by  \citet{wan10}.

\subsection{Limits on the mass-loss rate from hot subdwarfs}
\label{sec_ul}

The non-detection  of accretion-powered  X-ray emission in binaries containing compact objects (Sect.~\ref{sec_ulsdb}) can be used to constrain  the properties of stellar winds from the hot subdwarfs in these systems.  Thanks to their higher X-ray  efficiency, binaries containing NSs can provide more stringent limits than those containing WDs. Although  no confirmed NS+sd system has been found yet, a few candidates  which deserve further study have been proposed \citep{gei10}. 
We note that the constraints on the stellar mass-loss derived in this way are affected by a few  uncertainties, such as  the source distances, the X-ray spectrum assumed to convert count rates to fluxes,  the accretion model and geometry used to relate the X-ray luminosity to the mass accretion rate.  
 With these caveats in mind, we can discuss the results obtained for the WD+sdB binary \cd\  \citep{mer14} and for a dozen of other  sdB binaries which are suspected to host compact objects \citep{mer11b}. 
 
Under the assumption of Bondi-Hoyle accretion (eq. (1)), and assuming a power law spectrum with photon index $\Gamma$=2,  a 3$\sigma$ upper limit  of   $\mdot\sim3\times10^{-13}$ $\msun$ yr$^{-1}$ was derived for the mass-loss rate of \cd . The distance of this source is well known (364 pc, \citet{gei13}), thus the main uncertainty is related to the assumed X-ray spectrum.  As discussed in \citet{mer14}, any thermal spectrum with temperature in the range 0.1-10 keV, would give  a limit on $\mdot$  in the range  $(2-4)\times10^{-13}$ $\msun$ yr$^{-1}$.  This value is inconsistent with the theoretical predictions of \citet{vin02} if \cd\ has solar or higher abundances (see  Fig.~\ref{fig_vc02}). 

In Fig.~\ref{fig_vc02} we show also the upper limits on $\mdot$   obtained  from the  X-ray observations  of  three sdB binaries   which might have   NS companions  \citep{mer11b}.  These limits are  quite constraining,  but subject to the effective presence of a NS in these systems, which is based on the mass limits derived under the assumption that the subdwarf rotates synchronously. Further observations to confirm this are required. 
On the other hand, the current upper limits derived for the other sdB stars with proposed WD companions are two or three orders of magnitude higher and thus not particularly constraining for the theoretical models.

\begin{figure}
\begin{center}
\hbox{
\includegraphics[height=9cm,angle=-90]{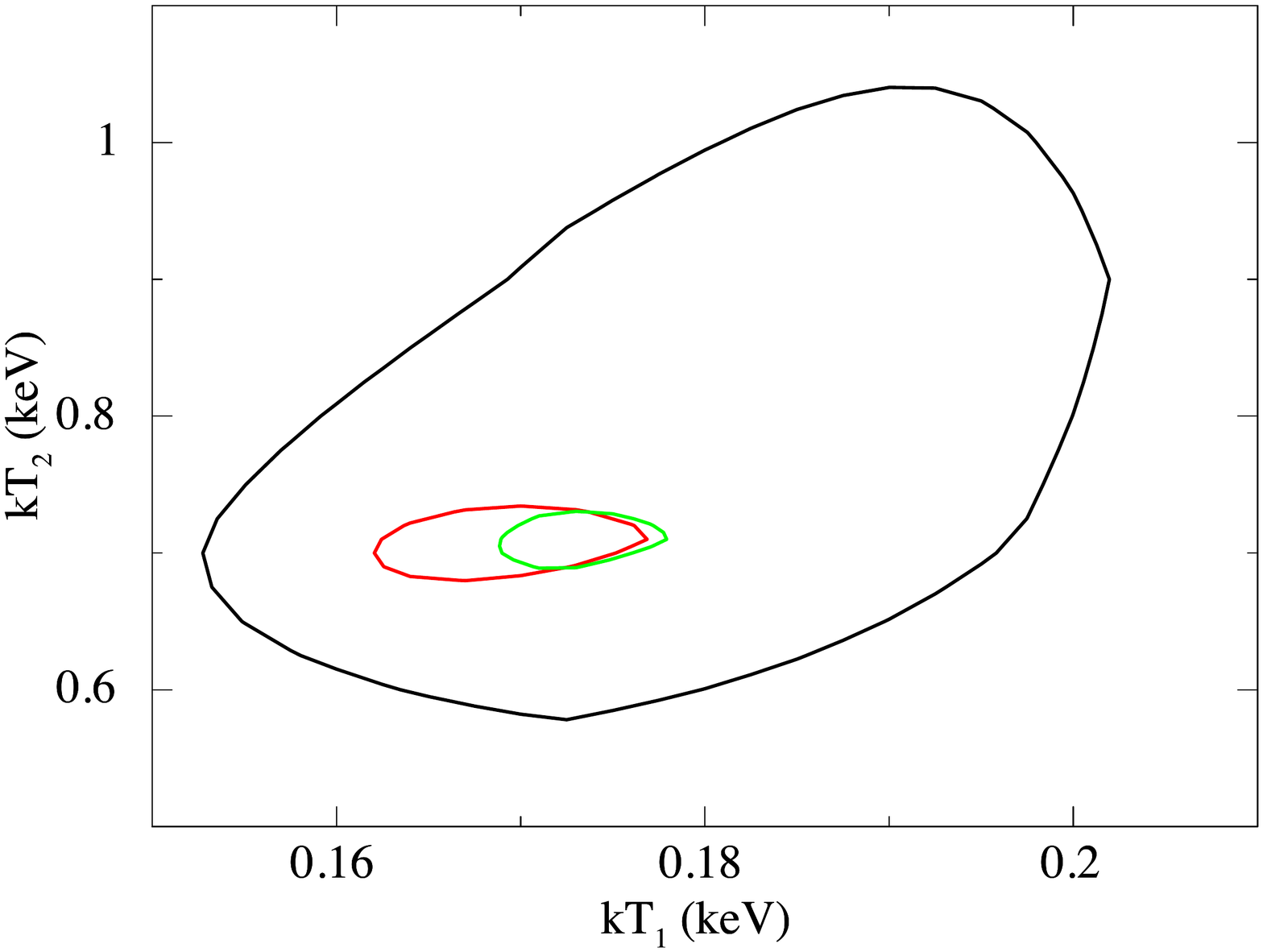}
\hspace{-2cm}
\includegraphics[height=9cm,angle=-90]{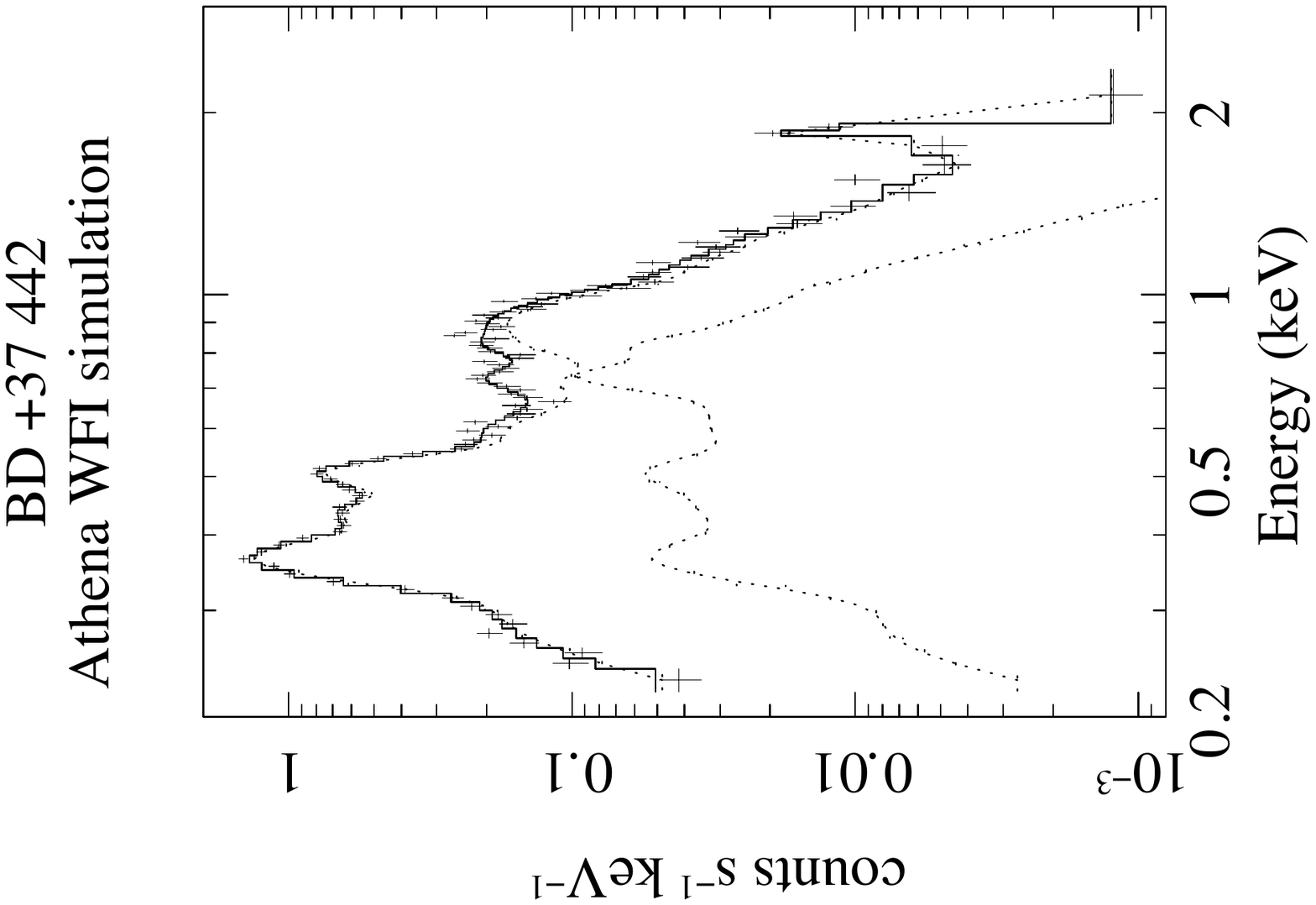}
}
\end{center}
\caption{Left panel: Contour plots showing the uncertainty on the temperatures of the 
two thermal components in the X-ray spectrum of \bd . The black line indicates the current error region (2$\sigma$ confidence level)
obtained with \xmm\ data. The red and green lines indicate the error regions derived from a simulated observation (50 ks exposure time) with the Athena  WFI and X-IFU instruments, respectively. 
Right panel: Simulated   spectrum of \bd\ for a 50-ks long observation with the Athena/WFI instrument.
\label{fig_simu}}
\end{figure}

\section{Conclusions and future prospects}

The results obtained for the small, but rapidly growing, sample of hot subdwarfs
detected in the X-ray range, show the potential of X-ray observations in the study of these objects.
Unfortunately, detailed  analysis such as those carried out for several normal OB stars cannot be done  for these
relatively faint sources. For example,  the element abundances of the  thermal plasma  models used to fit the  sdO spectra 
could not be constrained by the X-ray data currently available.   Therefore, the abundances had to be fixed to the values derived from optical studies, when available, or otherwise based on informed best guesses.

The increased sensitivity and spectral resolution of future X-ray telescopes  will provide much better 
information on the spectral properties of the  X-ray emission and lead to the discovery of many new sources. 
As an illustration of what could  be 
obtained with {\it Athena}, the next large  X-ray satellite approved by the  European Space Agency for a launch in the late 2020s,
we  have simulated  X-ray spectra of \bd\ based on  the expected performances of the   WFI and X-IFU instruments  \citep{rau14,bar14}.  
The analysis of these simulated data (see Fig.~\ref{fig_simu}) indicate that it will be possible to 
obtain much more accurate measures of the plasma temperatures and 
constrain within $\sim$10\% the abundances of C, N, O, Si and Fe, providing a measure independent 
of that derived from optical/UV analysis.

It is also important to increase the sample of hot subdwarfs detected in X-rays in order to consolidate the comparison 
with normal early type stars and to explore possible dependences on different parameters, 
such as surface gravity, temperature  and composition.  
While future satellites will certainly allow us to reach fainter limiting fluxes, also the current facilities
can in principle provide many new detections, since only a very small number of targets have been observed so far.
One interesting question is whether   isolated sdB stars are X-ray emitters.

More optical/UV studies of the X-ray emitting hot subdwarfs are needed to derive better estimates
of their atmospheric and wind parameters.   The accurate distances which will   soon be provided by the {\it Gaia} mission will be crucial to reduce many of the uncertainties involved in these estimates.
A multi-wavelength approach can be very effective to discover other systems such   as \hd\  and \cd , in which the presence  of a compact companion is certain. Besides their intrinsic interest from the point of view of binary evolution models, such systems  will allow us to  fully exploit the X-ray data to better characterize the poorly known winds of hot subdwarfs.
 
 \medskip
 \noindent
 {\bf Acknowledgements}
 \smallskip
 
 We thank the organizers of  the Science Event  ``X-ray Astrophysics of Hot Massive Stars'' at the 2014 COSPAR in Moscow and the participants to the Seventh Meeting on Hot Subdwarfs in Oxford for very interesting discussions. 
We are grateful to all our collaborators in the X-ray and optical observations of hot subdwarfs, and  in particular to A.Tiengo, P.Esposito, G.L.Israel, U. Heber and S.Geier.  This work has been  partially supported from PRIN INAF 2014. 
  



\end{document}